\documentclass[prl,aps,twocolumn,superscriptaddress]{revtex4}
\usepackage{graphicx}
\usepackage{dcolumn}
\usepackage{times}
\usepackage{amssymb,amsfonts,amsmath}
\usepackage{xr}
\usepackage[]{units}
\usepackage[]{nicefrac}
\usepackage{mathrsfs}
\usepackage{amsmath}

\begin{document}



\title{Volume changes during active shape fluctuations in cells} 
\author{Alessandro Taloni}
\affiliation{CNR - Consiglio Nazionale delle Ricerche,  Istituto per l'Energetica e le Interfasi,
Via R. Cozzi 53, 20125 Milano, Italy}
\affiliation{Center for Complexity \& Biosystems and 
Department of Physics, University of Milano, via Celoria 16, 20133 Milano, Italy}
\author{Elena Kardash}
\affiliation{Departments of Biochemistry and Molecular Biology, Sciences II, 30 Quai Ernest-Ansermet, CH-1211 Geneva 4, Switzerland}
\author{Oguz Umut Salman}
\affiliation{CNR - Consiglio Nazionale delle Ricerche,  Istituto per l'Energetica e le Interfasi,
Via R. Cozzi 53, 20125 Milano, Italy}
\affiliation{CNRS, LSPM UPR3407, Universit\'e Paris 13, Sorbonne Paris Cit\'e, 93430 Villetaneuse, France}
\author{Lev Truskinovsky}
\affiliation{LMS, CNRS-UMR 7649, Ecole Polytechnique, 
  Route de Saclay, 91128 Palaiseau, France}
\author{Stefano Zapperi}\email{stefano.zapperi@unimi.it}
\affiliation{Center for Complexity \& Biosystems and
Department of Physics, University of Milano, via Celoria 16, 20133 Milano, Italy}
\affiliation{CNR - Consiglio Nazionale delle Ricerche,  Istituto per l'Energetica e le Interfasi,
Via R. Cozzi 53, 20125 Milano, Italy}
\affiliation{Institute for Scientific Interchange Foundation, Via Alassio 11/C, 10126 Torino}
\affiliation{Department of Applied Physics, Aalto University,
P.O. Box 14100, FIN-00076, Aalto, Finland}
\author{Caterina A. M. La Porta}\email{caterina.laporta@unimi.it}
\affiliation{Center for Complexity \& Biosystems and Department of Bioscience, University of Milano, via Celoria 26, 20133 Milano, Italy}


\begin{abstract} 
Cells modify their volume in response to changes in osmotic pressure but
it is usually assumed that other active shape variations do not involve significant volume fluctuations. Here we report experiments demonstrating that water transport in and out of the cell is needed for the formation of blebs, commonly observed protrusions in the plasma membrane driven by cortex contraction. We develop and simulate a model of fluid mediated membrane-cortex deformations and show that a permeable membrane is necessary for bleb formation which is otherwise impaired. Taken together our experimental and theoretical results emphasize the subtle balance between hydrodynamics and elasticity in actively driven cell morphological changes.
\end{abstract}
\maketitle

Cells can change their shape to explore their environment, communicate with other cells and self-propel.  These macroscopic changes are driven by the coordinated action of localized motors transforming chemical energy into motion. Active processes in biological systems can be linked to a large variety of collective non-equilibrium phenomena  such as phase-transitions, unconventional fluctuations, oscillations and pattern formation 
\cite{Marchetti2013,Ramaswamy2010,Manneville2001}. A vivid example of actively driven 
non-equilibrium shape fluctuations is provided by cellular blebs, the rounded membrane protrusions formed by the separation of the plasma membrane from the cortex as a result of acto-myosin contraction \cite{CharrasPaluch2008,Fackler2008,Paluch2013}. 

Blebs occur in various physiological conditions \cite{Fackler2008,Paluch2013}, as for instance during zebrafish 
embryogenesis \cite{Raz2003,Blaser2006,Richardson2010,Kardash2010,Tarbashevich2010}, or cancer invasion \cite{Fackler2008}. While some questions concerning the mechanisms governing bleb formation and its relation to migration have been resolved \cite{CharrasPaluch2008,Fackler2008,Paluch2013,Charras2005,Charras2008,Tinevez2009,Kardash2010}, key aspects of bleb mechanics remain unclear. Geometrical constraints dictate that active shape changes associated with blebs should necessarily involve either fluctuations in the membrane surface or in cellular volume, and possibly both. It is generally believed, however, that the cellular volume is not significantly altered during bleb formation, so that the cell is usually considered incompressible \cite{Charras2005,Blaser2006,Tinevez2009}. Yet, experimental evidence in vitro suggests that aquaporins (AQPs), a family of transmembrane water channel proteins \cite{King1996}, are involved in cell migration \cite{Verkman2005,Monzani2009,Stroka2014} and blebbing \cite{Huebert2010,Karlsson2013}. The implied significance of fluid transport through the membrane suggests that an interplay between hydrodynamic flow and active mechanics has an important but still unclear role in blebbing.  In this letter, we reveal the role of the membrane permeability in the formation, expansion and retraction of cellular blebs. We show by direct experiments in vivo and numerical simulations that bleb formation involves volume fluctuations, considerable water flow through the membrane, and relatively smaller surface fluctuations. 

{\it Experiment:} One of the limitations impeding the experimental studies of the bleb dynamics is the lack of proper tools to generate high-resolution spatial-temporal data of bleb dynamics. This is due to the fact that the time scale of bleb formation is relatively short (about 1 minute starting from initiation of the bleb to its retraction), which requires fast imaging and photostable markers. Here, we create an improved membrane marker \cite{Nguyen2005} which we inject in one-cell stage zebrafish embryos
(see Supplemental Material for experimental methods \cite{suppl}). Together with wild type (WT) zebrafish PGCs,  a well studied biological  model to investigate blebbing in vivo \cite{Raz2003,Blaser2006,Richardson2010,Kardash2010,Tarbashevich2010},  we also consider cells expressing dominant-negative Rho Kinase mutant (DN-ROK) which inhibits acto-myosin contractility suppressing blebbing activity \cite{Blaser2006}. Zebrafish expresses a large number AQPs contributing to the water permeability of
the membrane \cite{Tingaud-Sequeira2010}.  Here we focus on AQP1 and AQP3, the most ubiquitously expressed aquaporins \cite{Tingaud-Sequeira2010,suppl}. To asses their role in volume change and blebbing, we consider PGCs with AQP1 and AQP3 overexpression (AQP+) and knockdown (AQP-). 

\begin{figure}[htb] \centering 
\includegraphics[width=\columnwidth]{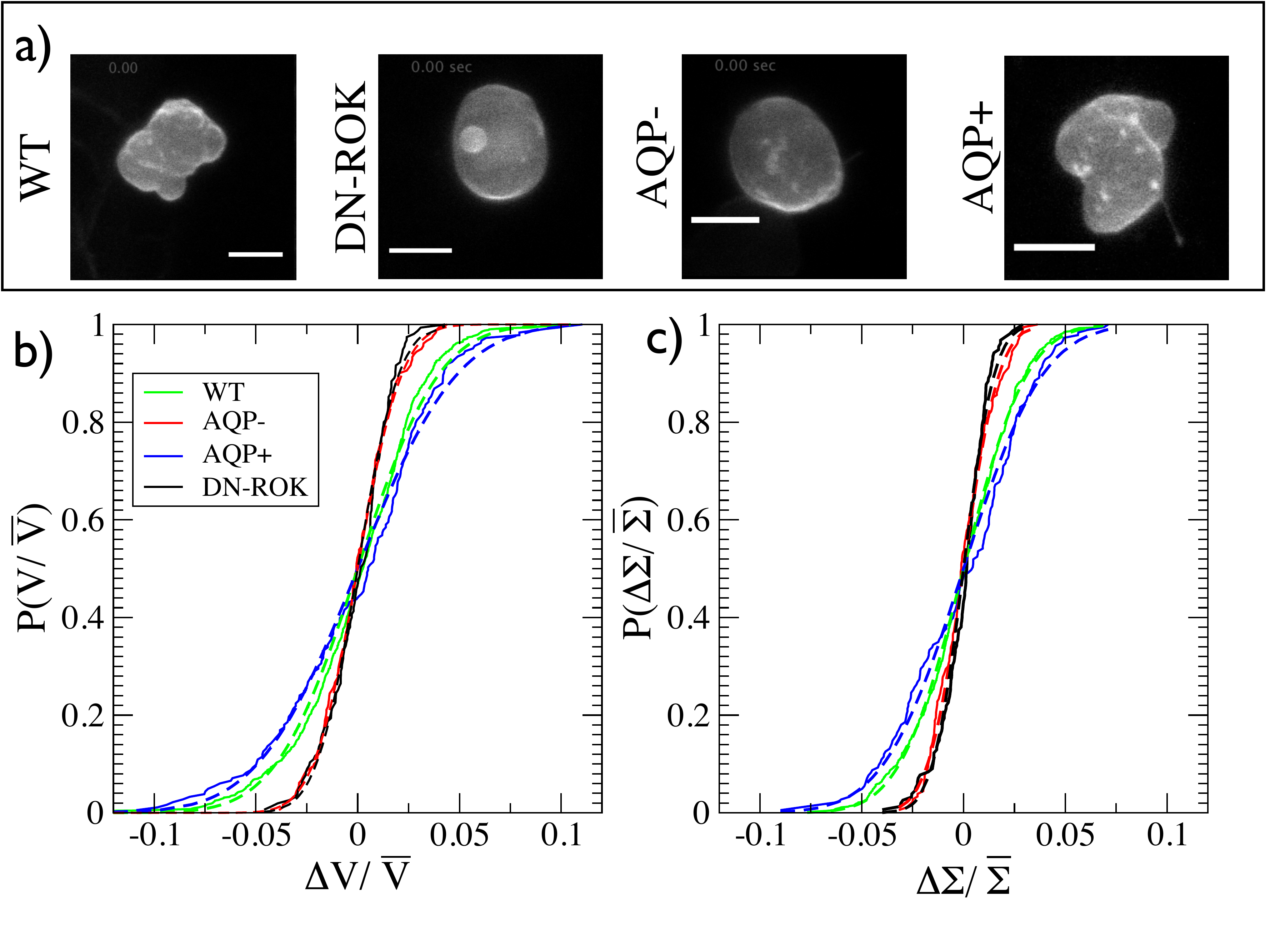}
 \caption{\label{fig:volume-surface} (Color online) Volume and surface fluctuations during blebbing are controlled by AQPs. a) Representative phenotypes for bleb formation in PGCs under different conditions: wild type (WT), cells expressing DN-ROK mutant which impairs contractility by interfering with acto-myosin contraction  (DN-ROK), cells where AQP1a and 3a is suppressed (AQP-), cells with over-expression of AQP1a/3a (AQP+). The scale  bar corresponds to $10 \mu \mathrm{m}$. The cumulative distribution of relative volume (b) and surface (c) fluctuations for the four conditions illustrated in a) together with a Gaussian fit (dashed lines). The distributions are sampled over $N$ different time-frames corresponding to $n$ different cells. (WT: $n=15$ cells and $N=800$ time-frames; DN-ROK: $n=2$ and $N=155$, AQP-:  $n=4$ and $N=180$; AQP+:  $n=7$ and $N=183$).}
\end{figure}

The imaging of blebbing is done during 12-16 hours post fertilization, when we record time-series of confocal images for a large number of cells. Sequences of image stacks are then processed using the 3D Active Meshes algorithm implemented in the Icy software \cite{Dufour2011}. The algorithm performs three-dimensional segmentation and tracking using a triangular mesh that is optimized using the original signal as a target. From the resulting three dimensional mesh one can then measure the cell volume and its surface area (see also \cite{suppl}). In Fig. \ref{fig:volume-surface}a, we illustrate representative phenotypes of  PGCs under different conditions (see also the corresponding movies \cite{suppl}). These observation show that WT cells display a marked blebbing activity that, as expected, is strongly suppressed  when active contraction is hindered, as in DN-ROK cells \cite{Blaser2006}. Remarkably, we also observe a strong reduction in bleb activity in AQP- cells, where water flow is hindered. In contrast,  water flow enhances blebbing as manifested by the presence of larger blebs in the AQP+ condition. To quantify these qualitative observations, we measure the cell volume $V$ and its surface $\Sigma$ sampling the results over a large number of time-frames taken on different cells. To account for cell-to-cell variability, we consider relative volumes $\Delta V/ \bar{V} = (V-\bar{V})/\bar{V}$ and surfaces $\Delta \Sigma/\bar{\Sigma} \equiv (\Sigma-\bar{\Sigma})/\bar{\Sigma}$ changes, where $\bar{V}$ ($\bar{\Sigma}$) is the time-averaged volume (surface) of each cell. 
The average value of the volume $V$ does not change significantly for the four cases \cite{suppl}.

\begin{figure}[ht] \centering 
\includegraphics[width=\columnwidth]{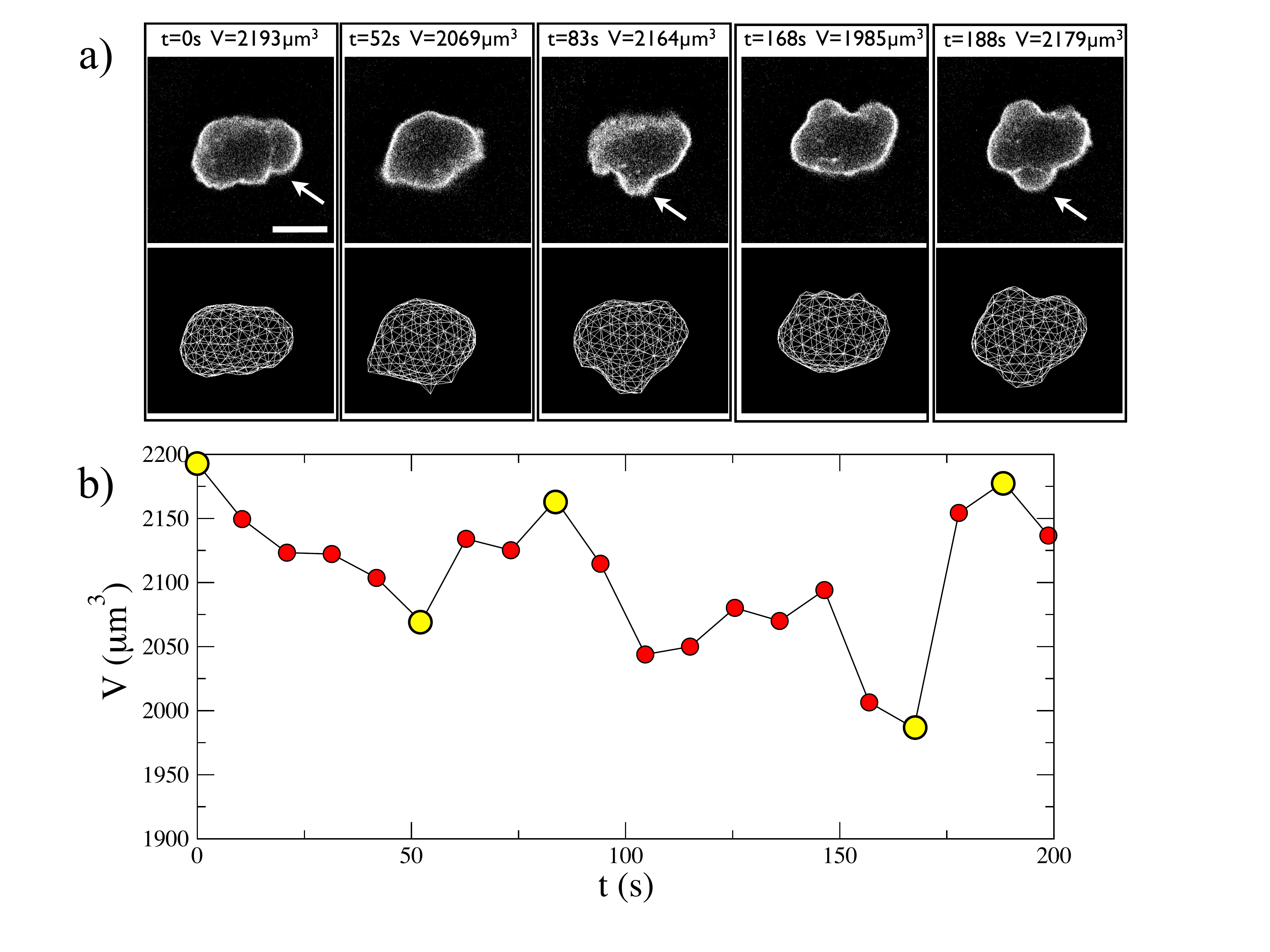}
 \caption{\label{fig:volume-bleb-time} (Color online) Bleb formation is directly related to volume changes.  a) Time lapse of a single AQP+ cell. Two dimensional sections are shown in the upper panels  and the corresponding reconstructed three-dimensional meshes in the lower ones.  Arrows
 indicate blebs. The scale bar corresponds to $10 \mu \mathrm{m}$. b) Evolution of the volume for
 the same cell. Time points corresponding to the panels in a) are denoted in yellow. An increase in volume is observed in correspondence with each newly formed bleb.} 
 \end{figure}

In Fig. \ref{fig:volume-surface}b, we report the cumulative distribution of relative volume changes which indicates significant fluctuations, reaching up to 10\%, in the WT case. Volume fluctuations are strongly reduced for the DN-ROK and AQP- cases, while they are  enhanced in the AQP+ case. The relative volume and surface distributions themselves are well described by Gaussian statistics, as also confirmed by a Kolmogorov-Smirnov test \cite{suppl}. A statistical test also indicates that the differences between WT and both AQP- and DN-ROK are significant ($p<0.01$), but the  differences between WT and AQP+ and between AQP- and DN-ROK are not \cite{suppl}. Relative surface fluctuations are 
small that volume fluctuations in the WT case and are further reduced for DN-ROK and AQP- and slightly increased for AQP+ (see Fig. \ref{fig:volume-surface}c). We also checked that relative surface and volume fluctuations are correlated \cite{suppl}, suggesting a direct link between blebbing activity and volume fluctuations induced by water transport. Suppressing water flow has the same effect as suppressing active contraction, in both cases blebs are hindered. Furthermore, the volume fluctuations we observe follow closely bleb expansion and retraction as illustrated in Fig. \ref{fig:volume-bleb-time}. Expansion of a bleb correspond to a visible volume increase while when a bleb retracts the volume decreases.  Here we concentrate on AQP+ PGC since the blebs are  distinctly visible and the analysis clearer, but the same result holds for WT cells where, however, several blebs may form and retract simultaneously. 

{\it Model:} In order to better understand the physical role of water flow in bleb formation, we 
resort to numerical simulations of a two-dimensional model of the biomechanics of
cortex-membrane deformations including fluid transport through the plasma membrane.
Several existing computational models for cellular blebs simulate the detachment and expansion 
of the membrane due to the active contraction of the cortex, assuming cell volume conservation \cite{Young2010,Lim2012,Lim2013,Strychalski2013}. Here, we relax this assumption
by introducing and varying the membrane permeability $\alpha$ \cite{Stockie2009} in a model based on the immersed boundary method \cite{Peskin2002} in the Stokes approximation \cite{Cortez2001}, considering a contracting discretized elastic cortex coupled to an elastic membrane. 
 

We describe both the membrane and the cortex by a set of discrete nodes connected by springs on  one dimensional closed curves parametrized by their initial arc length $s$ with $s\in[0,2\pi]$.  Using Lagrangian coordinates, the position of node $i$ is denoted by   $\mathbf r_{m_i}(t)$ (for the membrane) and $\mathbf r_{c_i}(t)$ (for the cortex).
Initial  positions  are  chosen to  be  $\mathbf r_{m_i}=(r_m \cos(s),r_m \sin(s))$ and $\mathbf r_{c_i}=(r_c \cos(s),r_c \sin(s))$ where $r_m$ and $r_c$ are the membrane and cortex radii, respectively.   
The interactions between the nodes on their respective curves include nearest-neighbor (NN)  and 
three-body interaction terms to account for stretching  and bending energies. We also model the cortex-membrane interface by a set of springs with random stiffness $k_{\mathrm{mc}}$ $(\unit{Nm^{-3})}$, drawn from a uniform distribution. Disorder
in the stiffness represents at a coarse-grained scale the random arrangement of cortex-membrane linker proteins. 
 The energy of the system can be decomposed as  $\mathscr E=\mathscr E_{\mathrm{c}}+\mathscr  E_{\mathrm{m}}+\mathscr  E_{\mathrm{int}}$.  The first two terms are given by
 \begin{eqnarray}
  \mathscr E_{\mathrm{x}} =\epsilon_\mathrm{x}\sum_{i=1}^{N} \biggl[\frac{k_\mathrm{x}}{2} \biggl(\frac{|\mathbf{r_{\mathrm{x}_{i+1,i}}}|-\epsilon_\mathrm{x}}{\epsilon_\mathrm{x}}\biggr)^2 + \\ \nonumber \frac{B_\mathrm{x}}{2\epsilon_\mathrm{x}^2}\biggl(\cos(\theta_{\mathrm{x}_{i-1,i,i+1}})-\cos(\theta^0_{\mathrm{x}_{i-1,i,i+1}})\biggr)^2 \biggr] \\ \nonumber
 \end{eqnarray}
where x corresponds to either membrane (m) or cortex (c),  $|\mathbf{r}_{\mathrm{x}_{i+1,i}}|$ is the distance between node $i$ and $i+1$, $\epsilon_\mathrm{x}$ is the equilibrium distance between NN nodes, $k_\mathrm{x}$ (N/m) is the stiffness coefficient, $B_\mathrm{x}$ (J) is the bending coefficient, $\theta_{{i-1,i,i+1}}$ and $\theta^0_{i-1,i,i+1}$ are the angles between the triplets $(i-1,i,i+1)$ in deformed  and equilibrium configurations, respectively. The interaction energy is given by 
$\mathscr E_{\mathrm{int}}=\sum_i k_{\mathrm{mc}} (|\mathbf r_{\mathrm{mc}_{i}}| -l)^2/2$, where
$|\mathbf{r}_{\mathrm{mc}_{i}}|$  and $l$ are the distances between the membrane and cortex node with the same index $i$ in the deformed and  equilibrium configurations. A non-vanishing rest length $\epsilon_\mathrm{c}$ for the cortex element is needed to prevent that the cortex collapses under hydrostatic forces and represents a convenient method \cite{Barnhart2010,Du2012,Recho2013} to account for osmotic regulation present in cells \cite{Jiang2013}. 
Cortex elements are assumed to follow over-damped dynamics
$\mu_c   \mathbf{\dot r}_{c_{i}}  = - \delta\mathscr E/\delta \mathbf{r}_{c_{i}}$,
where $\mu_c$ $(\unit {kgm^{-2}s^{-1}})$ is the cortical drag coefficient.

We first consider an impermeable elastic membrane that moves with the fluid velocity satisfying a no-slip boundary condition.   The fluid velocity $\mathbf u$ and the pressure $p$ satisfy  Stokes equation with the incompressibility constraint
\begin{equation}
\mu\Delta\textbf{u} =\nabla p-\textbf{f},  ~ ~ ~\nabla\cdot\textbf{u} = 0, \label{eq:velocity}
\end{equation}
where $\mu$ is the fluid viscosity and  $\textbf{f}$ is the body force per unit volume $\unit{(Nm^{-3}})$ that can be calculated by spreading the  force density $\mathbf F_{\mathrm{m}_{i}}$  from solid (Lagrangian $\mathbf r_{\mathrm{m}_{k}}$)  to fluid (Euler $\mathbf r$) coordinates as $\mathbf f(\mathbf r) = \sum_{k=1}^{N}\mathbf F_{\mathrm{m}_{k}} \delta_h(\mathbf r-\mathbf r_{\mathrm{m}_{k}})$,
where $\mathbf F_{\mathrm{m}_{k}}=- \delta\mathscr E/\delta \mathbf{r}_{m_{k}}$ and $\delta_h(\mathbf r)$ is the two dimensional discretized  delta function.
Eq. (\ref{eq:velocity}) is solved using the regularized Stokeslet method \cite{Cortez2001} as in Ref. \cite{Lim2013}.
After the velocity has been calculated in Euler coordinates, we evolve
the membrane nodes as $\dot{\mathbf r}_{\mathrm{m}_{k}}= u(\mathbf r_{\mathrm{m}_{k}})$.
We take special care to correct $u(\mathbf r_{\mathrm{m}_{k}})$ to enforce volume conservation, a common problem of the immersed boundary method \cite{suppl,Stockie2009,Newren2007}.

Cortex contraction and healing are the main driving forces of the blebbing activity. In our model, we assume that the cortex is pre-stretched which we impose by choosing a value for the equilibrium distance between the nodes $\epsilon_\mathrm{c}$ that is smaller than their initial distances. Bleb nucleation in our model occurs stochastically due to the randomness of the stiffness of the bonds representing the membrane-cortex interface. Local detachment is 
then implemented by setting to zero the bond stiffness $k_{\mathrm{mc}}$ if its stretching 
$|\mathbf{r}_{\mathrm{mc}_{i}}|$  is above a threshold that we set at $0.1 l$. Finally, to implement interface healing, we assume that each cortex node $i$, associated to a removed interface bond, moves towards the membrane
with velocity $\mathbf{v} = \nu_\mathrm{c} \mathbf{r}_{\mathrm{mc}_{i}}/|\mathbf{r}_{\mathrm{mc}_{i}}|$ until it attaches again when $|\mathbf{r}_{\mathrm{mc}_{i}}| \leq l$ \cite{Lim2013}. 
When disconnected nodes become connected again, we assign new random values to the
stiffness $k_{\mathrm{mc}}$ of the bond.  We observe that the strength of disorder (i.e. the wideness of its distribution) controls bleb nucleation which in turn may cause the presence of a large number of blebs, occurring simultaneously. As in previous models \cite{Young2010,Lim2012,Lim2013,Strychalski2013}, we neglect additional time-dependent effects due to viscoelasticity and actin turnover in the cortex \cite{Fritzsche2013}.

We introduce permeability into the model by following Ref. \cite{Stockie2009}. We assume Darcy Law and
impose a porous slip velocity normal to the membrane given by
 \begin{equation}
 u_p(\mathbf r_{\mathrm{m}_{k}})=-\frac{K}{\mu}\frac{\partial P}{\partial n}\approx-\frac{K}{\mu}\frac{[P]}{a},
 \end{equation}
where $K$ is permeability, $\mu$ is the viscosity, $a$ is membrane thickness and $[P]$ is the pressure jump.
Using the normal stress jump condition $[P]= \nicefrac{\mathbf {\mathbf F}\cdot \mathbf n }{\Delta S(\mathbf r_{\mathrm{m}_{k}})}$, the porous velocity $u_\mathrm{p}$ reads
$u_\mathrm{p}(\mathbf r_{\mathrm{m}_{k}})=-\alpha \mathbf {F}\cdot \mathbf n/\Delta S(\mathbf r_{\mathrm{m}_{k}})$
where $\alpha=K/(\mu a)$ $(\unit{m^2s kg^{-1}})$ and $\mathbf { F}$ is the force on the node. Finally, we reach the equation used in the simulations
$\dot{\mathbf r}_{\mathrm{m}_{k}}=  u^{\mathrm{corr}}(\mathbf r_{\mathrm{m}_{k}}) + u_\mathrm{p}(\mathbf r_{m_{k}})$.

{\it Simulations:} To simulate the model, we assume a square fluid domain that we discretize using a square
grid with a discretization step $dx=L/N_\mathrm{E}$ and $dy=L/N_\mathrm{E}$, where $L$ is the length of the domain in each direction and $N_\mathrm{E}\times N_\mathrm{E}$ is the number of Eulerian coordinates. The fluid domain covers both the inside and outside of the membrane and has an area of $100$ $\times100\unit{\mu m}$. The discretization steps $dx$ and $dy$ are of size $1$ $\unit{\mu m}$.

\begin{figure}[ht] \centering 
\includegraphics[width=\columnwidth]{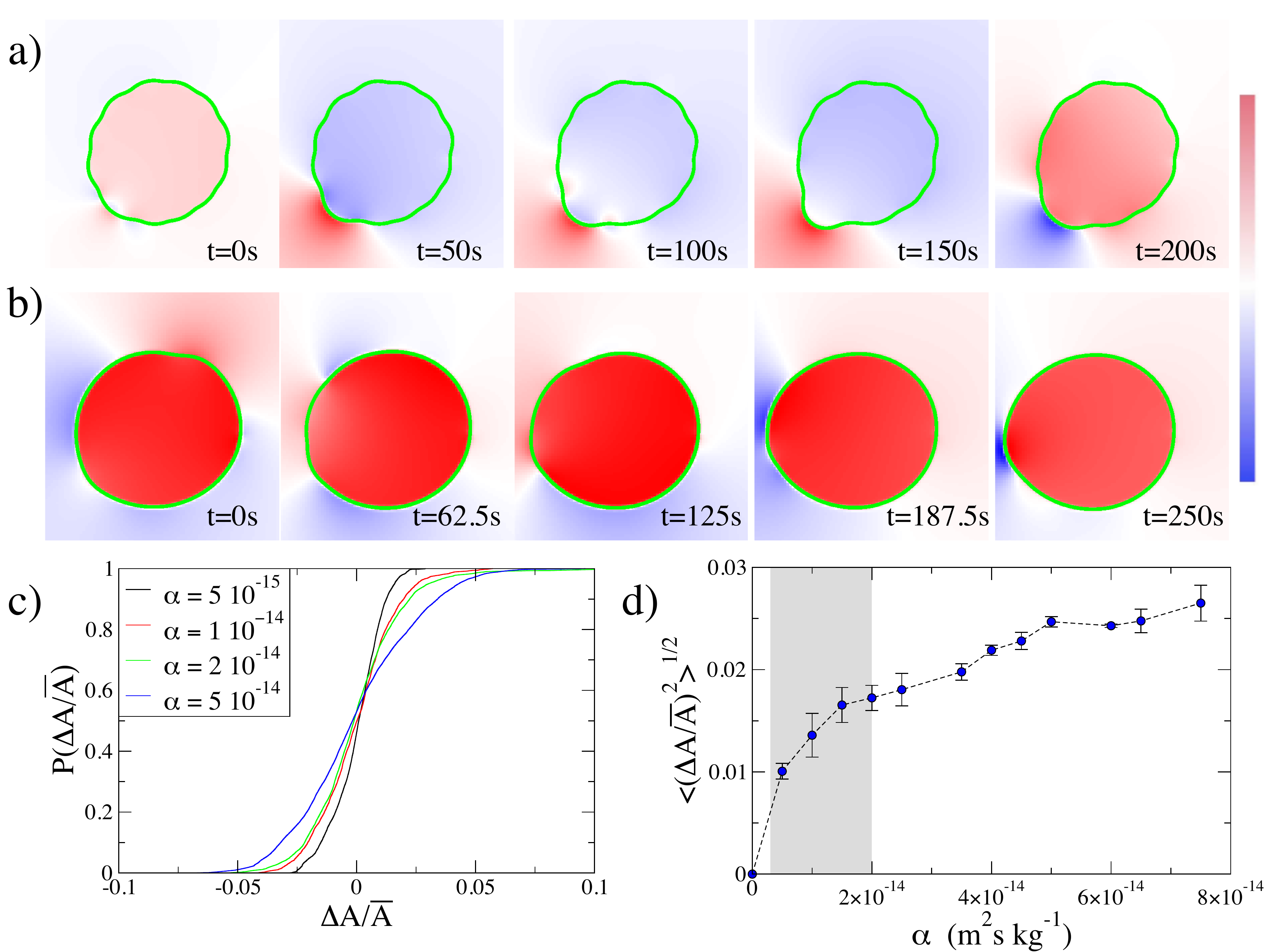}
 \caption{\label{fig:simulations} (Color online) Numerical simulations show that membrane porosity is needed for blebbing. a) A series of snapshots of the numerical simulations of bleb formation and retraction. The color represents the local fluid pressure: red for positive and blue for negative pressures while the cell membrane is green. a) Results obtained with physiological permeability: $\alpha=2 \cdot 10^{-14}\mathrm{m}^2\mathrm{s/kg}$.  b)  For $\alpha=0$, bleb formation is impaired. c) The cumulative distribution of relative area fluctuations for different permeabilities. d) The standard deviation of  the distribution of relative cell areas as a function of permeability. The physiological range is depicted in grey.}
 \end{figure}

The membrane permeability in zebrafish embryos has been measured experimentally and
is reported to be in the range $3\times 10^{-15}<\alpha < 2\times 10^{-14} \mathrm{m^2 s/kg}$ depending on the developmental stage \cite{Hagedorn1997}. Here, we perform numerical simulations under different values of $\alpha$, ranging from $\alpha=0$ to $\alpha = 8 \cdot 10^{-14} \mathrm{m^2 s/kg}$, to account for AQP overexpression and knockdown (see Table S1 for a complete list of parameters). When $\alpha$ is in the physiological range, we observe realistic bleb formation and retraction (Fig. \ref{fig:simulations}a and Movie S5). 
When we delete membrane permeability, setting $\alpha=0$ and enforcing strict cell volume conservation, bleb activity is suppressed (Fig. \ref{fig:simulations}b and Movie S6). We can relate the simulations to the experimental results by noticing that the cell area $A$, the two dimensional analogue of the three dimensional cell volume $V$,  fluctuates more or less when the permeability is increased or reduced (Fig. \ref{fig:simulations}cd), in correspondence with AQPs over-expression or knock-down, respectively \cite{Tingaud-Sequeira2010}. Both experiments and simulations suggest that blebs occur as long as the fluid is able to flow sufficiently fast through the membrane. 

\begin{figure}[ht] \centering 
\includegraphics[width=\columnwidth]{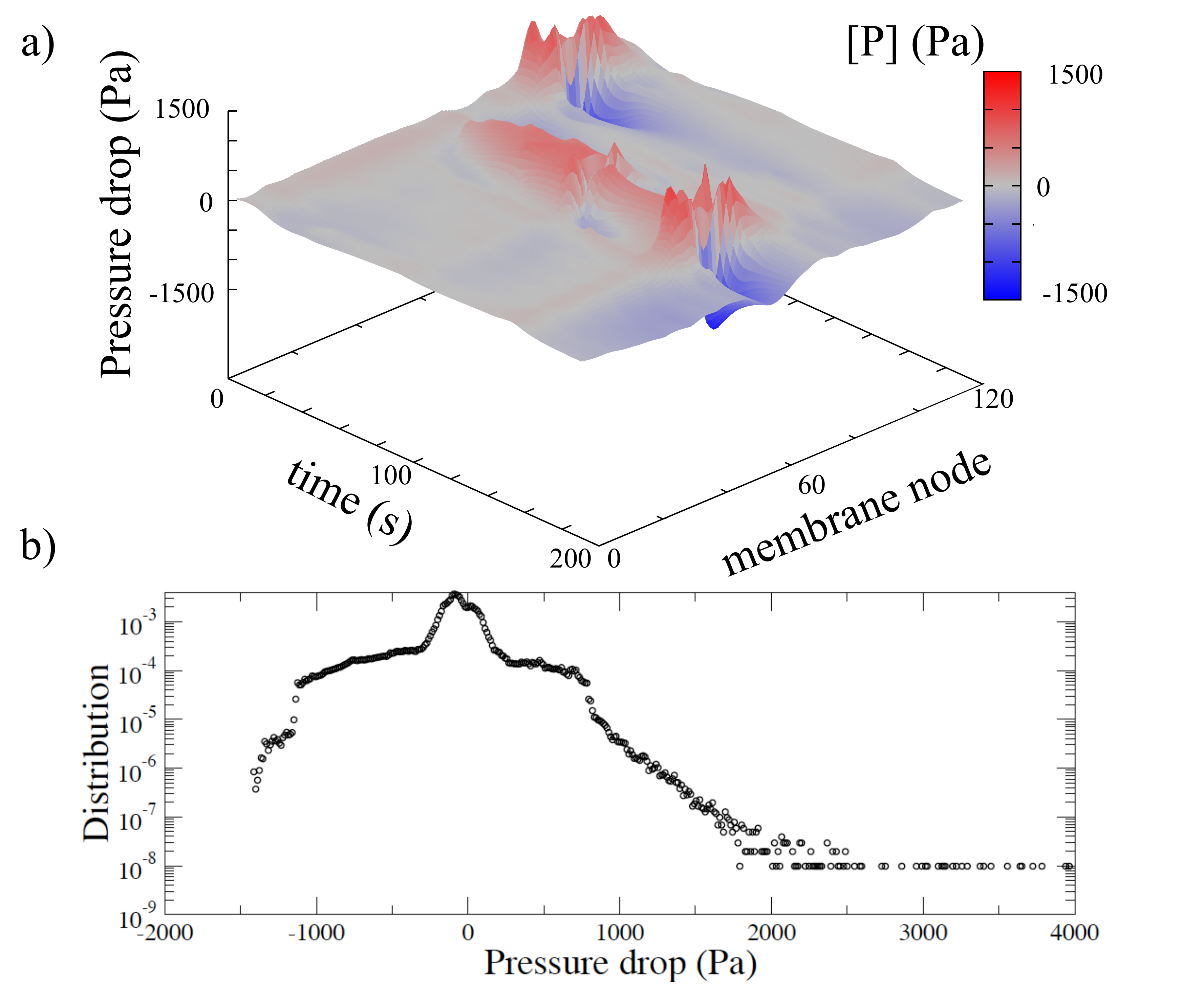}
 \caption{\label{fig:drop} (Color online) a) Spatio-temporal evolution of the pressure jump across the membrane
computed from the normal stress jump condition  
in numerical simulations with $\alpha=2 \cdot 10^{-14}\mathrm{m}^2\mathrm{s/kg}$. The pressure drop is strongly
 enhanced in localized regions corresponding to blebs. b) The corresponding 
 distribution of pressure jumps displays long tails.}
 \end{figure}

{\it Discussion:} Our model allows us to better understand why volume fluctuations are crucial for blebbing. Acto-myosin driven cortex contraction leads to a shrinkage of the cell by squeezing some water outside.  Indeed the fluid pressure inside the cell is initially larger than the one outside (see Fig. \ref{fig:simulations}a). The contraction of the cortex induces stretching in the membrane-cortex linkers leading the membrane to buckle \cite{Katifori2009}. Buckling provides an effective way for membranes to avoid considerable elastic compression and is associated with a structural softening of the system \cite{Kierfeld2006,Sens2007}. Furthermore, it allows to generate large membrane deflections needed to form a bleb. In differentiated cells additional membrane surface can be obtained by disassembling caveolae \cite{Sinha2011}, but this can not happen in PGCs where caveolin is not expressed. When the interface fractures, the mechanical stress on the detached part of the membrane decreases, but it increases on the interface that is still attached, inducing crack propagation. As the bleb expands, the fluid pressure inside the cell is reduced  (see Fig. \ref{fig:simulations}a) leading to an inflow of water. In Fig. \ref{fig:drop}a, we display the spatio-temporal evolution of the pressure jump across the membrane, showing large fluctuations in correspondence to bleb formation \cite{suppl}. These pressure spikes, whose distribution is long-tailed (Fig. \ref{fig:drop}b), are a manifestation of the stress concentrations around cracks and are needed to account for the observed volume fluctuations, because otherwise the average pressure jump generated by a uniform cortex contraction (around $10^2$ Pa) would not displace sufficient amount of fluid during the short lifetime of a bleb \cite{Tinevez2009,Jiang2013}. Healing of the membrane-cortex interface eventually leads to bleb retraction and to an increased fluid pressure inside the cell. The mechanism described above does not work for an impermeable membrane: 
isochoric buckling is possible in principle but, in addition to bending, it necessarily causes considerable stretching which is energetically expensive \cite{Katifori2009}. Thus, when the interface fractures the bleb does not form and interface delamination takes place without localized membrane expansion.

Our numerical results show that volume fluctuations during blebbing are due to mechanically induced highly
non-uniform pressure drops across the membrane. The same mechanism could also explain the experimental observation that
blebs are nucleated preferentially in regions of negative membrane curvature \cite{Tyson2014}. 
Changes in osmotic pressure gradients could also contribute to the process, as suggested previously \cite{Charras2008,Karlsson2013,You1996}, but are not explicitly included in our model. Future experimental
work will clarify if this and other assumptions present in our model are correct, but our results 
should stimulate both new experiments and the development of more elaborate theories and models.
This would also allow to better understand the role of transmembrane water transport for other cellular protrusions, given that past experimental results relate the presence of aquaporins to the formation of lamellipodia \cite{Monzani2009} and filopodia \cite{Karlsson2013}. The present methodology provides the basis for a physical explanation of this broad class of phenomena.

\begin{acknowledgments}

We thank A. Dufour, A. L. Sellerio and D. Vilone for useful discussions. 
A. T., O. U. S., and S. Z. are supported by the European Research Council through the Advanced
Grant No. 291002 SIZEFFECTS. S. Z. acknowledges support from the Academy of Finland FiDiPro progam, Project
No. 13282993. C. A. M. L. P. acknowledges financial support from MIUR through PRIN 2010. E. K. acknowledges
the support of a FEBS long-term postdoctoral fellowship while writing this manuscript.

A. Taloni, E. Kardash and O. U. Salman contributed equally to this paper.
\end{acknowledgments}


\section{Supplementary Informations}
\subsection{Zebrafish maintenance}
Zebrafish (Danio Rerio) of the AB and AB/TL genetic background were maintained, raised and staged as previously described \cite{Westerfield1993,Kimmel1995}. 

\subsection{Constructs and mRNA synthesis}
For protein over-expression in germ cells, the mRNA was injected into the yolk at one-cell stage. Capped sense RNA was synthesized with the mMessage mMachine kit (Ambion,\\ http://www.ambion.com/index.html). To direct protein expression to PGCs, the corresponding open reading frames (ORFs) were fused upstream to the 3’UTR of the nanos1 (nos1-3’UTR) gene, facilitating translation and stabilization of the RNA in these cells \cite{Koprunner2001}. 
For global protein expression, the respective ORFs were cloned into the pSP64TS ector that contains the 5’ and 3’ UTRs of the Xenopus Globin gene. The injected RNA amounts are as provided below. The following constructs were used:
\begin{itemize}
\item[-]YPet-YPet-RasCAAX-nos-1 (240 pg.) was used to label membrane in germ cells. 
\item[-] Lifeact-pRuby-nos-1 (240 pg.) was used to label actin in germ cells.  
\item[-]DN-ROK-nos-1 – (300 pg.) was used to interfere with ROK function in PGCs
\item[-]Aqp1a-nos-1 – (300 pg.) was used to over-express aquaporin-1a in PGCs
\item[-] Aqp3a-nos-1 – (300 pg.) was used to over-express aquaporin-3a in PGCs
\item[-]Aqp1a—EGFP-nos-1 – (360 pg.) was used to visualize the subcellular localization of aquaporin1a in PGCs 
\item[-]Aqp3a—EGFP-nos-1 – (300pg.) was used to visualize the subcellular localization of aquaporin3a in PGCs 
\end{itemize}

\subsection{Morpholino knockdown}

The morpholinos for knocking down protein translation were obtained from GeneTools, LLC http://www.gene-tools.com/. The following sequences were used: Aquapoin1a: 5’ AAGCCTTGCTCTTCAGCTCGTTCAT3’ (injected at 400$\mu$M);  Aquaporin 3a: 5’ ACGCTTTTCTGCCAACCCATCTTTC 3’ (injected at 400$\mu$M);. For the control, standard morpholino 5’CCTCTTACCTCAGTTACAATTTATA 3’ was used.

\subsection{Live Imaging of germ cells in zebrafish embryos}

Time-lapse movies of blebbing cells in live zebrafish embryos were acquired with the Zeiss LSM710 bi-photon microscope using one-photon mode. The 20x water-dipping objective with the numerical aperture 1.0 was used. The bit depth used was 16 and the scanning speed ranged between 150 to 250 ms/frame for fast imaging of bleb formation. 

\subsection{Image processing}

Images were preprocessed with Fiji software to eliminate the background. The Bleach Correction tool (EMBL) was used to correct for the reduction in fluorescence intensity during prolonged time-lapse movies.  Sequences of of image stacks were then processed using the 3D Active Meshes algorithm \cite{Dufour2011} implemented in the Icy software http://icy.bioimageanalysis.org/. The algorithm performs three-dimensional segmentation and tracking using a triangular mesh that is optimized using the original signal as a target. From the resulting three dimensional mesh one can then measure the cell volume its surface area. Three dimensional rendering of the meshes was done using Paraview (http://www.paraview.org/).

\subsection{Statistical analysis}
Statistical significance was evaluated using Kolmogorov-Smirnov tests implemented in custom made python codes. 

\subsection{Test of the accuracy of cell volume evaluations}

To confirm that the volume fluctuations observed in the zebrafish cells in vivo are real, we have to exclude 
possible systematic errors induced by the three dimensional mesh reconstruction algorithm \citep{Dufour2011}.
 Our experimental  analysis shows  smaller volume fluctuations for cells where the bleb formation has been suppressed (DN-ROK and AQP- mutants), with respect to wild type cells (WT) or those for which blebbing has been enhanced (AQP+). In particular, as shown in Fig. 1(a), WT and AQP+ cells display a more complex morphology when compared to DN-ROK or AQP-, which appear instead to have a rounded shape. Hence, the first question to be answered is whether in presence of complex morphological shapes, the algorithm introduces a systematic bias in the measured cell volume. In other words the question is: does
 the algorithm produce a larger error while calculating the volume of blebbing cells, with respect to those where blebs are absent? 

To answer to this question we generate a set of synthetic ellipsoidal cells whose volume is in the range of the zebrafish cells analyzed in our experiments ($\sim 2000-3000 \mu m^3$), as shown in Fig.\ref{fig:Synthetic_blebs}. Synthetic cells are generated through the ImageJ software (http://imagej.nih.gov/ij/) by  creating 3D stacks having the shape of ellipsoids with semi-axes $a_x, \, a_y$ and $a_z$. Initially, we set the voxel size to $0.13 \mu m$ along the three directions (see Fig.\ref{fig:Synthetic_blebs}(a)). The volume is then calculated both according to the formula $V_{ell}=4\pi/3 a_x a_y a_z$ and by counting the number of voxels belonging to the synthetic cell, $V_{wp}$. The relative error between the
two estimates is $\langle\frac{V_{wp}-V_{ell}}{V_{ell}}\rangle\simeq -0.05\%$, so that we can safely consider the estimate $V_{wp}$ as the real volume of the synthetic stacks generated. The main sources of error in analyzing confocal image
stacks stems from the anisotropic voxel. In our experiments the resolution along the $z$ direction is $0.78 \mu m$, while
it is  $0.13 \mu m$ in the xy plane. To reproduce the voxel anisotropy in the synthetic stacks, we select just one single xy plane every 6 composing the original z-stack. A resulting typical cell is shown in Fig.\ref{fig:Synthetic_blebs}(b).
We then extract the mesh of this newly obtained stack (see Fig.\ref{fig:Synthetic_blebs}(c)) and calculate the ensuing volume $V_{no-bleb}$.  Our set of synthetic cells consists of 35 ellipsoids of different semi-axes, for which we calculate the true volume $V_{wp}$  reported in Fig.\ref{fig:Volume_ellipsoids_comparison}. Then, each ellipsoid is first processed by the anysotropic voxelization in the z direction, and subsequently analyzed by the 3D Active Mesh algorithm. The  volumes of the extracted meshes are reported in Fig.\ref{fig:Volume_ellipsoids_comparison}  ($V_{no-bleb}$). It is apparent that  the algorithm systematically  underestimates the volume by roughly  $\delta=\langle \frac{V_{no-bleb}-V_{wp}}{V_{wp}}\rangle\simeq -14\%$. We checked that is error is greatly reduced for isotropic voxels.

A constant systematic error is not worrying, since we are only interested in changes in volume and all the images have the same voxel anisotropy and therefore the same error. Before addressing the fluctuations of this error, we focus on possible spurious changes in the measured volume induced by a change in shape. For each of the synthetic ellipsoidal cells, we create a synthetic cell with the same volume but presenting 1, 2, or 3 blebs on the surface. Blebs are generated as spherical caps of different radii with centers placed randomly on the ellipsoid surface (see Fig.\ref{fig:Synthetic_blebs}(d)). We then perform the same anysotropic voxelization of the original z-stack done for the plane ellipsoids (see Fig.\ref{fig:Synthetic_blebs}(e)). From this image we extract the active mesh  (Fig.\ref{fig:Synthetic_blebs}(f)) and calculate its volume. The volumes of the meshes of synthetic cells  with blebs, $V_{bleb}$, are displayed in Fig.\ref{fig:Volume_ellipsoids_comparison}. One can only see a very small difference between the values of $V_{bleb}$ and $V_{no-bleb}$, but they both appear underestimate the true value $V_{wp}$ by about $\delta\simeq -14\%$. What is surprising, however, is that cells with blebs appear to approximate the real volume slightly better than cells without blebs.  To confirm this,  we report in Fig.\ref{fig:volume-error-vsS} the relative volume fluctuations 
$\frac{V_{bleb}-V_{no-bleb}}{V_{no-bleb}}$ as a function of the measured mesh surface  fluctuations $\frac{\Sigma_{bleb}-\Sigma_{no-bleb}}{\Sigma_{no-bleb}}$. If no errors were made by the algorithm in estimating the synthetic volumes one would expect $\frac{V_{bleb}-V_{no-bleb}}{V_{no-bleb}}=0$ since pair of cells were constructed with the same volume but different shapes. If complex shape with blebs would lead to an overestimation of the volume with respect an ellipsoidal cell with no blebs, one would expect $\frac{V_{bleb}-V_{no-bleb}}{V_{no-bleb}}$  and $\frac{\Sigma_{bleb}-\Sigma_{no-bleb}}{\Sigma_{no-bleb}}$ to be positively correlated. To the contrary, the linear regression of the data in Fig.\ref{fig:volume-error-vsS} shows a small but clear anti-correlation between volume and surface fluctuations. This is in contrast with experimental results
showing that changes in shape are positively correlated with changes in volume (Fig. \ref{fig:VS_correlations}). Hence, this result can not be considered an artefact of the measurement but a real feature of the cells.

The previous analysis clearly demonstrates that the volume fluctuations observed in our experiments do not depend on the shape of the cells, but the algorithm systematically underestimates the volume of both of about $\delta\simeq -14\%$. The next question is whether these systematic fluctuations are of the same order of magnitude as the observed ones. Are the fluctuations reported in Fig. 1(b) real or just an artefact introduced by the mesh reconstruction algorithm? This question is particularly compelling in the case of WT and AQP+ cells, since for DN-ROK and AQP- cells we can accept that the volume might remain constant.
To answer to this question generate a set of 120 synthetic cells, 60 with blebs randomly placed and of different sizes, and 60 without blebs, each cell having its own real volume $V_{wp}$ calculated with ImageJ. We then process each synthetic cell according to the protocol previously outlined: anysotropic voxelization in the z direction and subsequent mesh analysis. Finally we calculate the fluctuations  $\frac{\Delta V}{\overline{V}}=\frac{V_{(no-)bleb}-V_{wp}}{V_{wp}}$ and its cumulative distribution $P\left(\frac{\Delta V}{\overline{V}}\right)$. In Fig.\ref{fig:Cumulative_fluctuation_ellipsoids_comparison_cell} we compare $P\left(\frac{\Delta V}{\overline{V}}\right)$ with the corresponding cumulative distributions of WT, AQP+, DN-ROK and AQP- cells, once it has been shifted by the average systematic volume bias $\delta \simeq -14\%$. This figure shows that systematic errors made by Icy in the estimation of the cell volume, are compatible with the observed volume fluctuations of DN-ROK and AQP- cells, but not with those AQP+ and WT cells, which instead appear to be significantly larger. We thus conclude that the 
difference in volume fluctuation between AQP+/WT cells and DN-ROK/AQP- is not an artefact of the analysis.

\subsection{Volume conservation in numerical simulation} 
By performing numerical simulations, we notice that 
a straightforward formulation of the method suffers from poor volume conservation. This general drawback of the immersed boundary method has been already pointed out by many other authors in different contexts \cite{Stockie2009,Newren2007}.  To overcome this problem we implement the method proposed in Ref. \cite{Newren2007} and enforce the incompressibility constraint on the discrete Lagrangian grid in the weak sense:
\begin{equation}\sum_{k=1}^{N}\mathbf u(\mathbf r_{\mathrm{m}_{k}})\cdot \mathbf n (\mathbf r_{\mathrm{m}_{k}})\Delta S(\mathbf r_{\mathrm{m}_{k}})=0,
\end{equation}
where $\mathbf n (\mathbf r_{\mathrm{m}_{k}})$ is the  outward unit normal to the membrane at the position 
$\mathbf r_{\mathrm{m}_{k}}$ and $\Delta S(\mathbf r_{\mathrm{m}_{k}})$ is a discrete measure of the arclength in the actual configuration  at the position $\mathbf r_{m_{k}}$. The  above constraint is satisfied by adding a corrective term to the equation of motion
 $\dot{\mathbf r}_{\mathrm{m}_{k}}=u^{\mathrm{corr}}(\mathbf r_{\mathrm{m}_{k}})$,
where $u^{\mathrm{corr}}\equiv u(\mathbf r_{\mathrm{m}_{k}}) - M\mathbf n (\mathbf r_{\mathrm{m}_{k}})$ satisfies the incompressibility constraint and  $M$ is given by 
 \begin{equation}
M=\frac{1}{\sum_{i=1}^{N} \Delta S(\mathbf r_{\mathrm{m}_{i}})}\sum_{k=1}^{N} \mathbf u(\mathbf r_{\mathrm{m}_{k}})\cdot \mathbf n (\mathbf r_{\mathrm{m}_{k}})\Delta S(\mathbf r_{\mathrm{m}_{k}}).
 \end{equation}
With this correction, we observe that the incompressibility constraint is satisfied
and the cell volume is perfectly conserved.

\section{Supplemental figures}

\renewcommand{\figurename}{Figure}
\setcounter{figure}{0}

\renewcommand{\thefigure}{S\arabic{figure}} 

 \renewcommand{\tablename}{Table}
\setcounter{table}{0}
\renewcommand{\thetable}{S\arabic{table}} 

\begin{figure}[h] 
\includegraphics[width=\columnwidth]{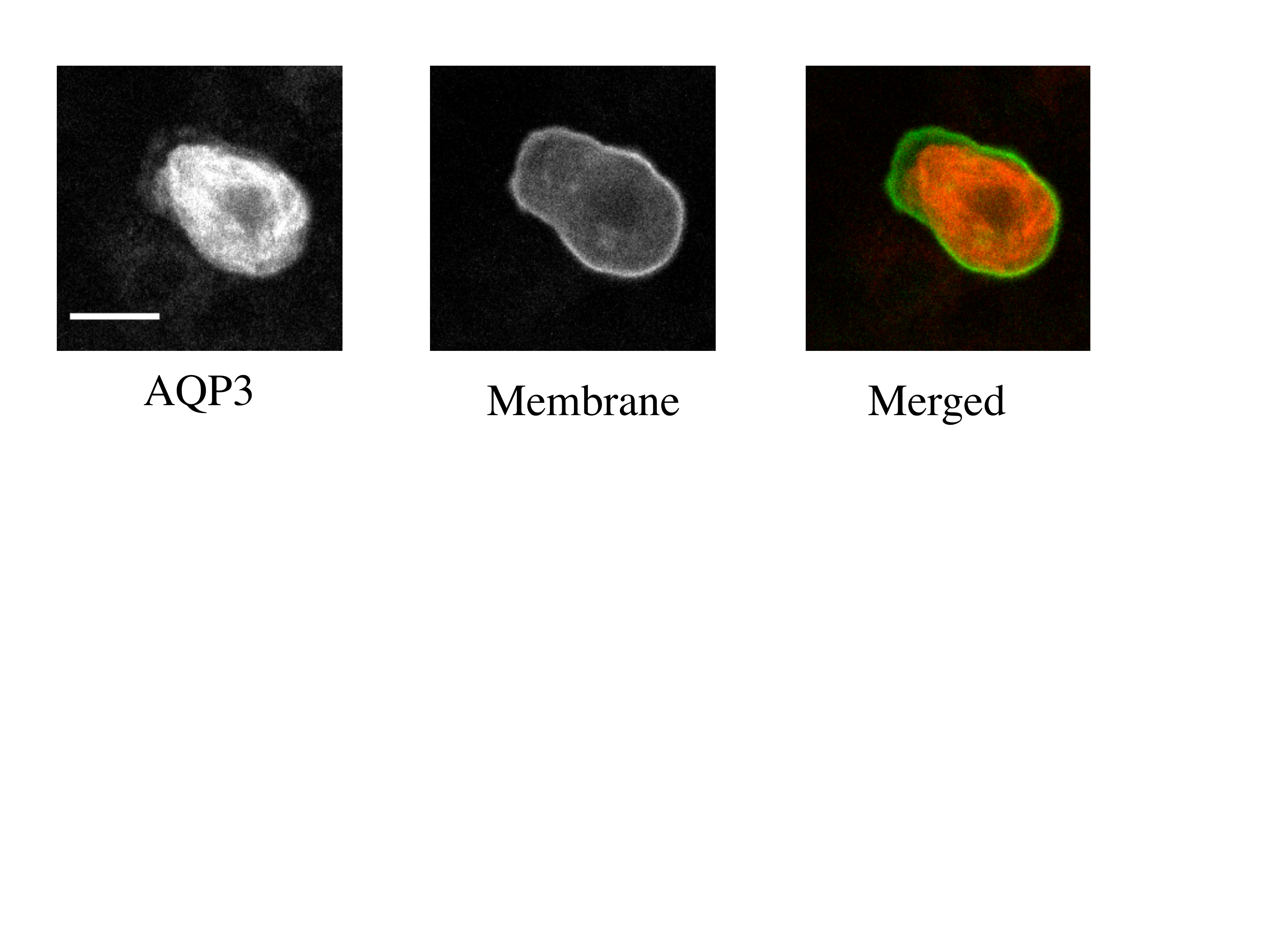}
 \caption{\label{fig:AQP_membrane} AQP3 is expressed by PGCs. A 
 PGC expressing EGFP fusion of Aquaporin-3a and an RFP-tagged membrane marker. 
 Scale bar is 10$\mu$m.}
 \end{figure}
  
  \begin{figure}[h] 
\includegraphics[width=\columnwidth]{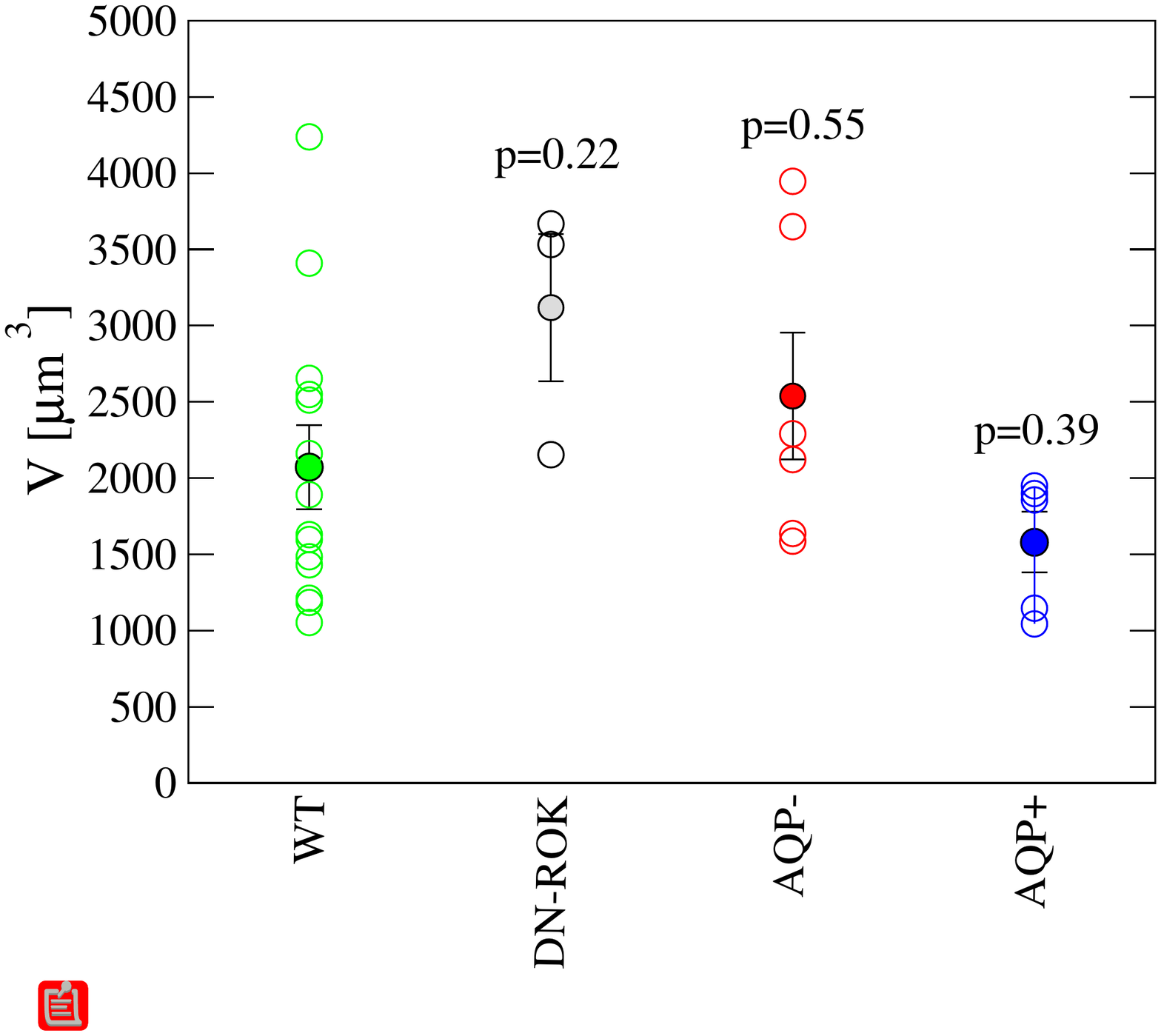}
 \caption{\label{fig:ABS_vol} Aquaporin knockdown and overexpression does not induce significant changes
 in the average cellular volume. A similar result holds for the DN-ROK mutant. The results show the dispersion
 of the data, the average and the standard error. Statistical significance ($p$ values) 
 is evaluated according to the Kolmogorov-Smirnov test. }
 \end{figure}

\begin{figure}[h] 
\includegraphics[width=\columnwidth]{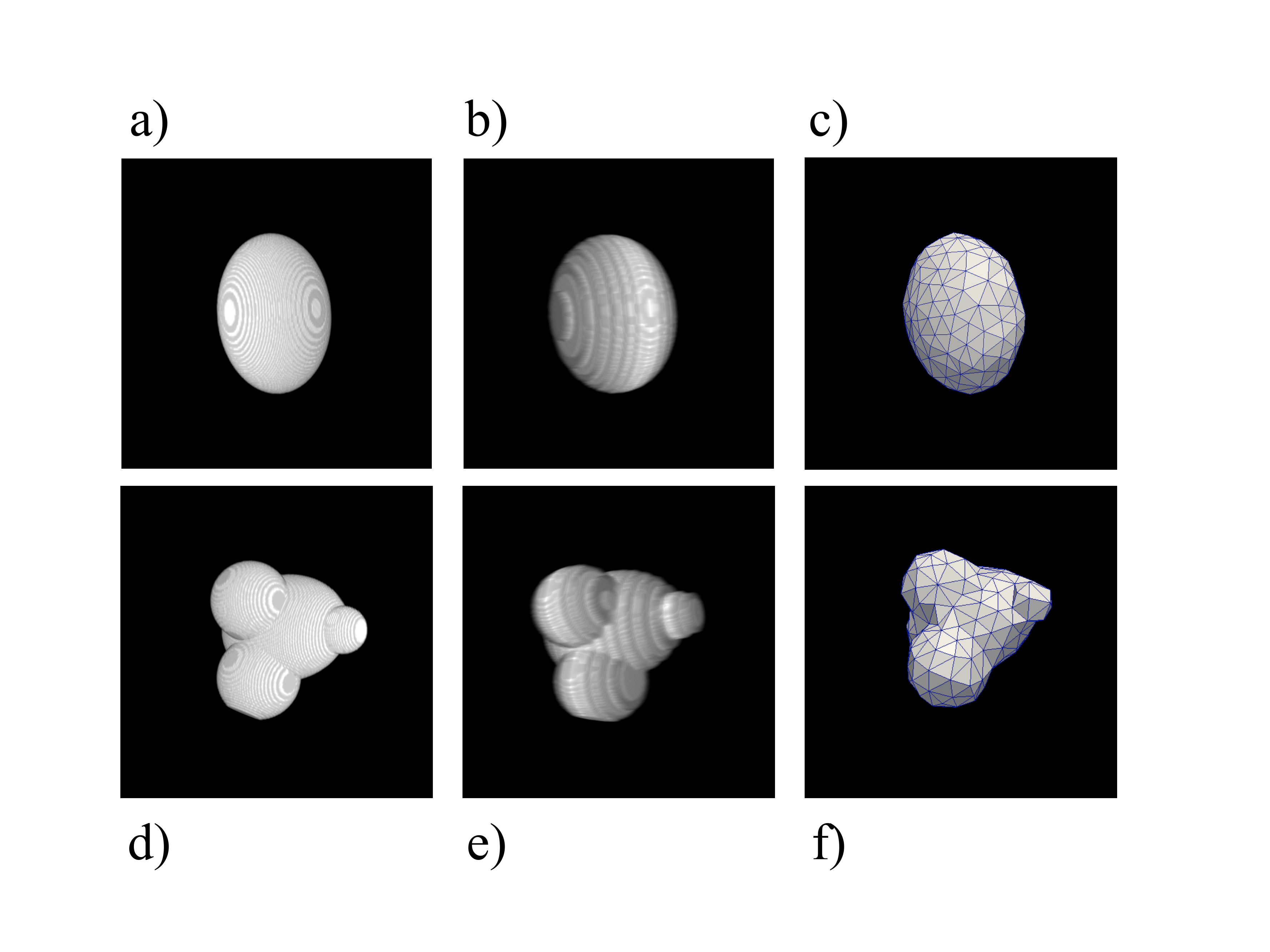}
 \caption{\label{fig:Synthetic_blebs} Two example of synthetic image stacks representing cells without (a-c) and
 with (d-f) blebs. The original stacks  of equal volume (a and d), are first transformed removing a set of planes (1 every 6) to obtain an anisotropic
 voxel corresponding to the experimental resolution (b and e), i.e. $0.13\mu$m in the x,y directions and $0.78 \mu$m along z. The resulting stacks (b and e) are analyzed to obtain a three-dimensional
 mesh (shown in c and f)}
 \end{figure}

\begin{figure}[h] 
\includegraphics[width=\columnwidth]{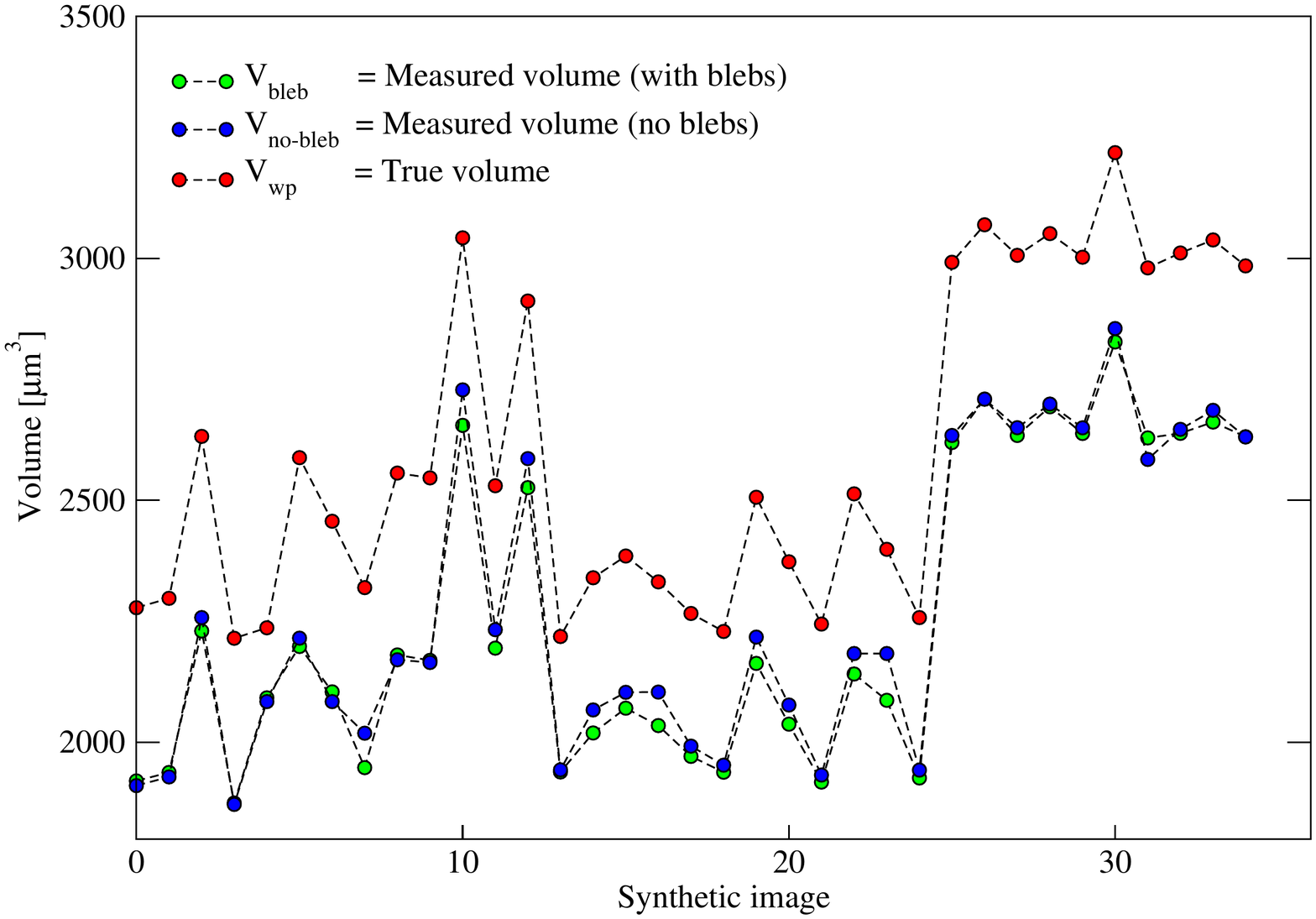}
 \caption{\label{fig:Volume_ellipsoids_comparison} Volume comparison of synthetic cells. We compare the measured volumes of cells without blebs ($V_{no-bleb}$) with those corresponding to cells with blebs  ($V_{bleb}$) and with the true volumes ($V_{wp}$), which is the same
 for each pair of cells. The results show a large constant systematic error $\delta\simeq -14\%$,  and small fluctuations for cells with and 
 without blebs.}
 \end{figure}

\begin{figure}[h] 
\includegraphics[width=\columnwidth]{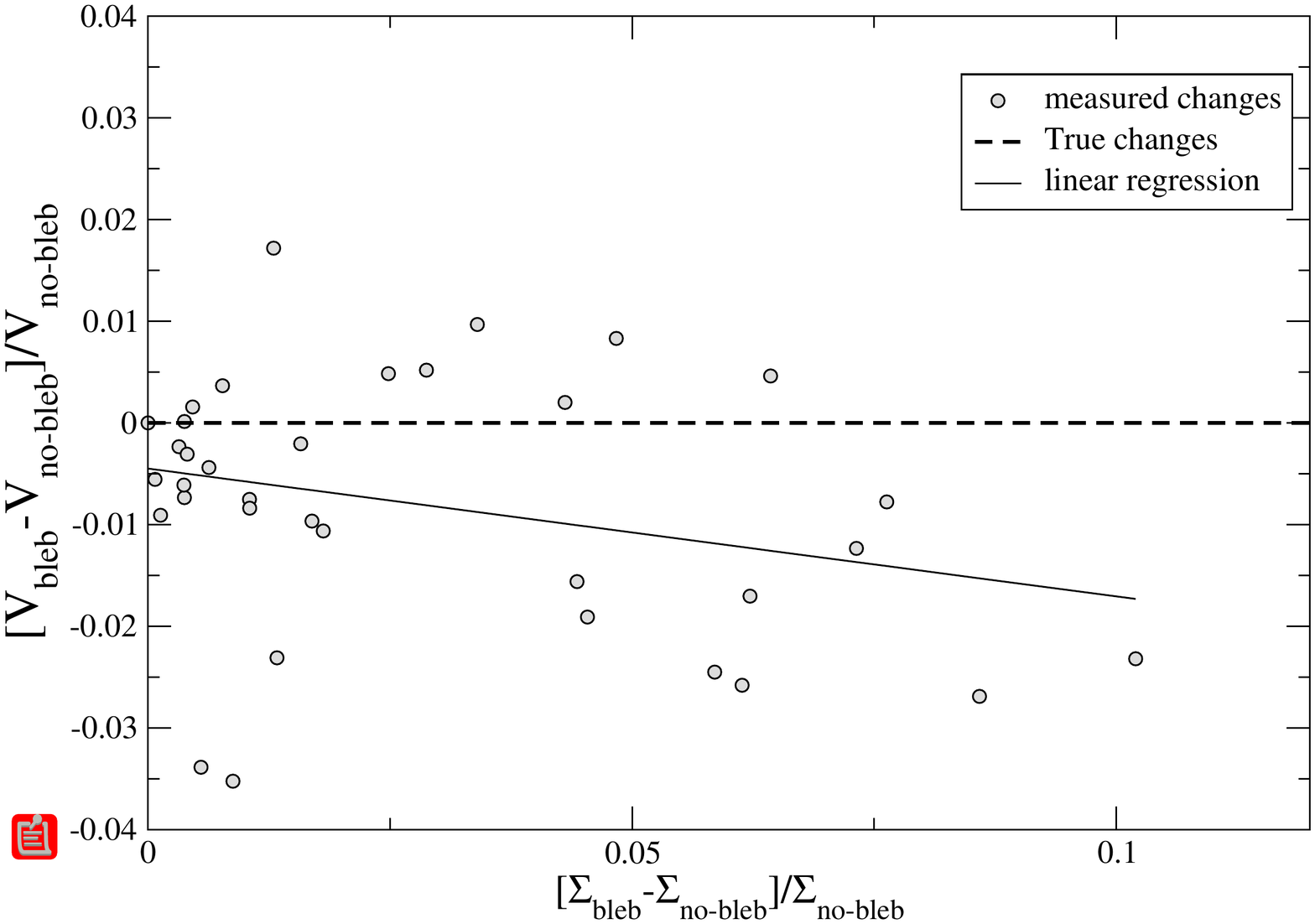}
 \caption{\label{fig:volume-error-vsS} The relative difference in the measured volume for pairs of cells of the same
 true volume but different surface due to the presence or absence of blebs. The data show fluctuations of $1\%$ and a 
 small negative correlation between volume and surface changes.}
 \end{figure}

\begin{figure}[h] 
\includegraphics[width=\columnwidth]{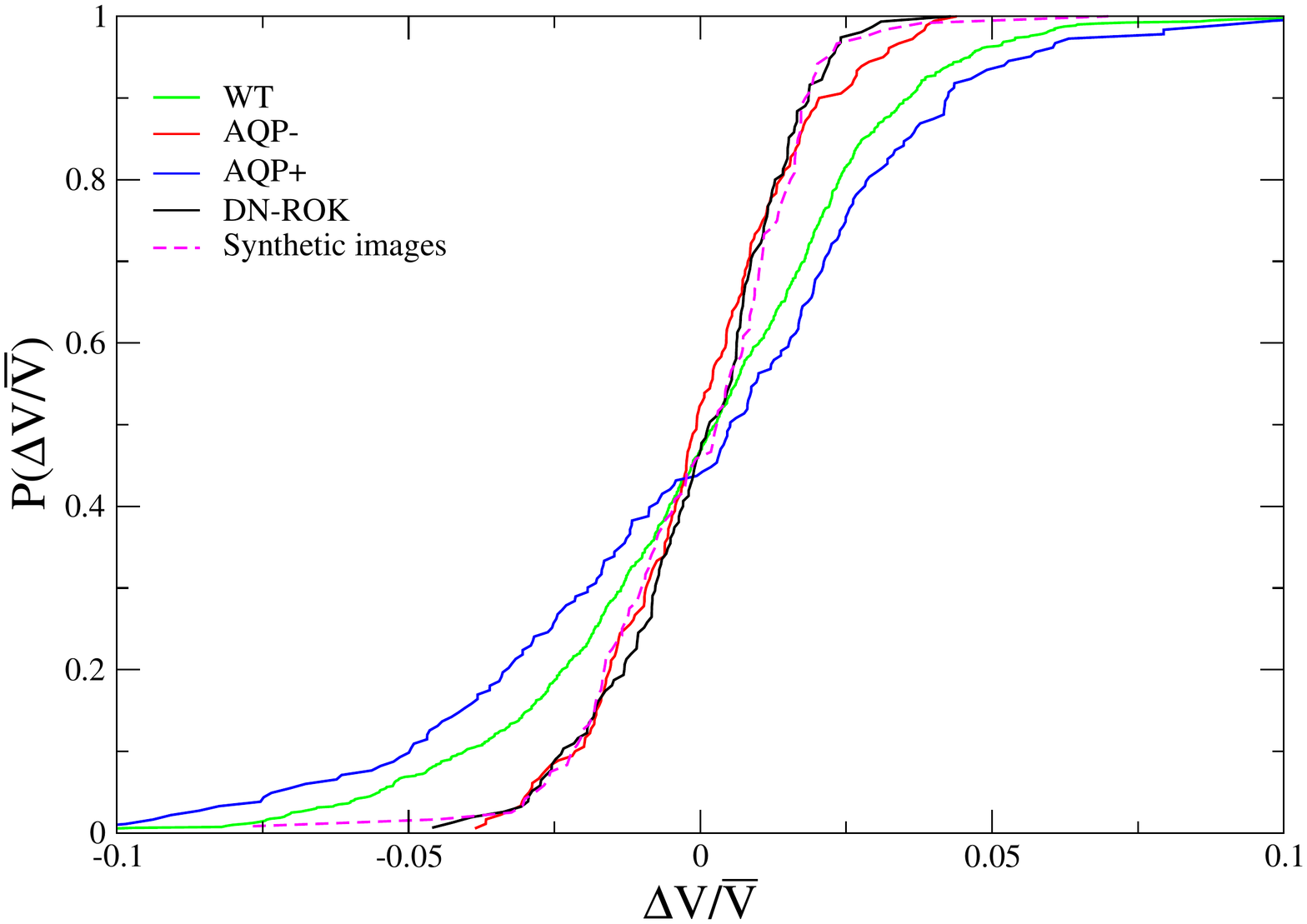}
 \caption{\label{fig:Cumulative_fluctuation_ellipsoids_comparison_cell} The cumulative distribution of relative error fluctuations for the volume of a large number of synthetic cells with and without blebs compared with experimental measurements. The cumulative distribution of synthetic cells is shifted by $-\delta$ to allow a visual comparison with the experimental quantities. $\overline{V}$ corresponds the time averaged volume for AQP+, AQP-, DN-ROK and WT cells, whilst it is $\overline{V}=V_{wp}$ for synthetic cells. Volume fluctuations for WT and AQP+ cells are significantly larger than those observed in synthetic cells, whereas DN-ROK and AQP- mutant cells volume fluctuations seem to be compatible with the systematic errors induced by the mesh algorithm.}
 \end{figure}

\begin{figure}[h] 
\includegraphics[width=\columnwidth]{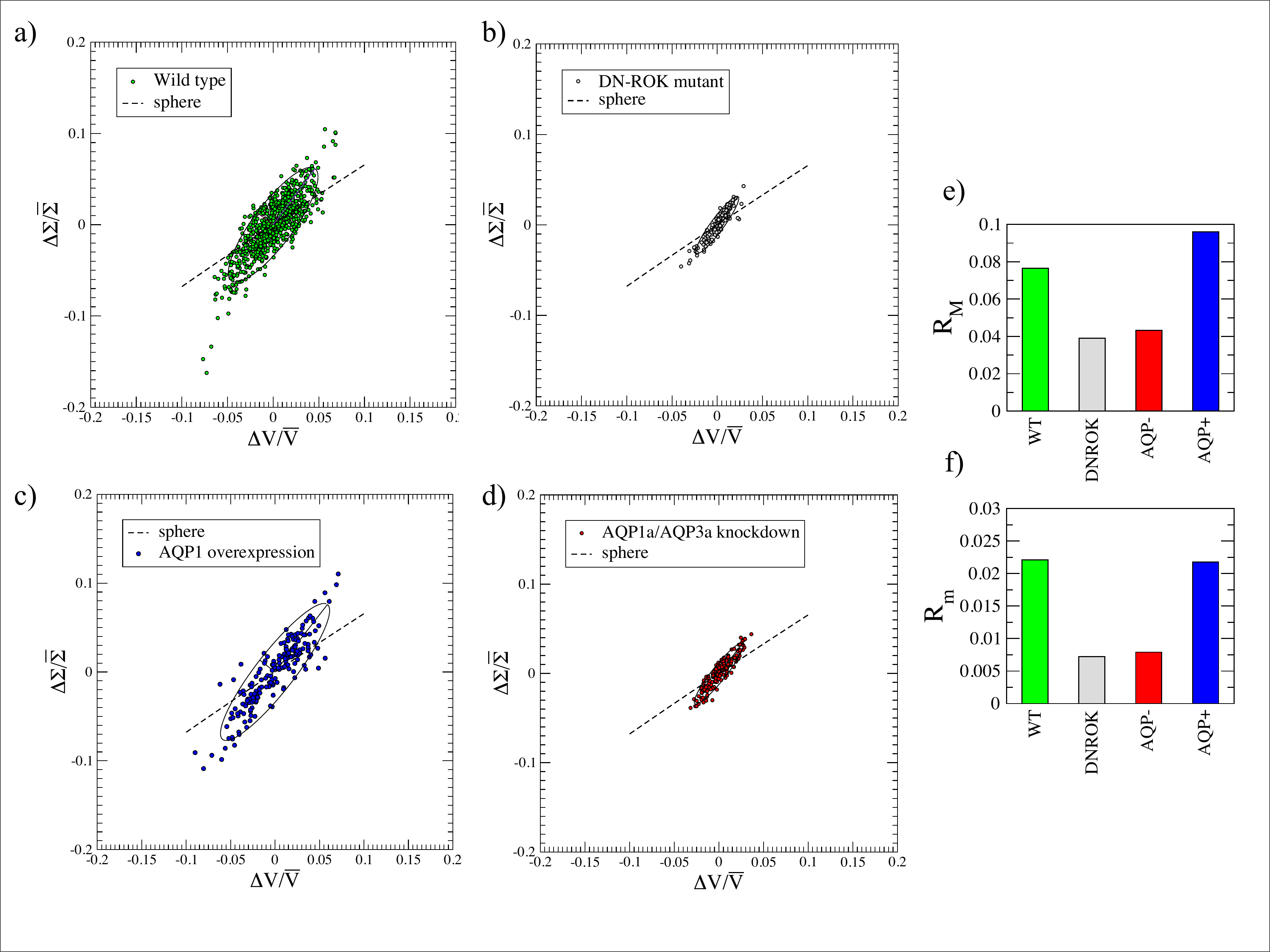}
 \caption{\label{fig:VS_correlations} Volume and surface fluctuations are correlated. Principal component
analysis of volume and surface relative values. Scatter plots for a) WT, b) DN-ROK,
c) AQP+ and d) AQP- are reported together with an ellipse with axis given by the
two eigenvectors of the cross-correlation matrix, whose amplitude is reported
in panel e) and f) for the largest and smaller axis. The dashed line represents
the expected result for the ideal case of the isotropic deformation of a sphere. 
  }
 \end{figure}

\begin{figure}[p] 
\includegraphics[width=\columnwidth]{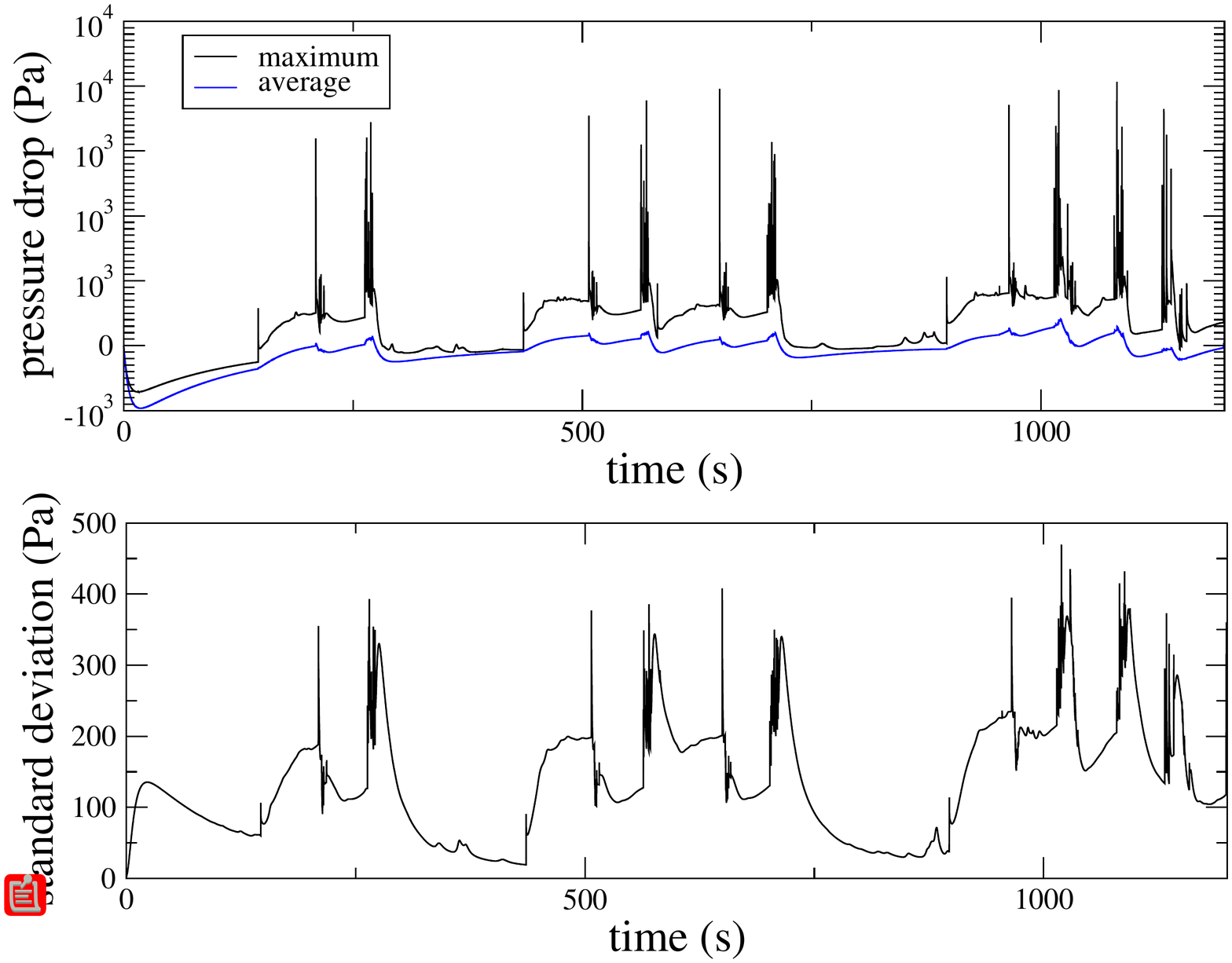}
 \caption{\label{fig:pdrop} Time evolution of the pressure drop across the membrane
from numerical simulations.A representative
 example of the evolution of the pressure drop in numerical simulations shows that the
 average is very different from the maximum  (top). Furthermore, the standard deviation of the distribution fluctuates intermittently in time in correspondence to the blebbing activity (bottom). Results are obtained for a permeability $\alpha=2 \cdot 10^{-14}\mathrm{m}^2\mathrm{s/kg}$.}
 \end{figure}

\begin{figure}[ht] \centering 
\includegraphics[width=\columnwidth]{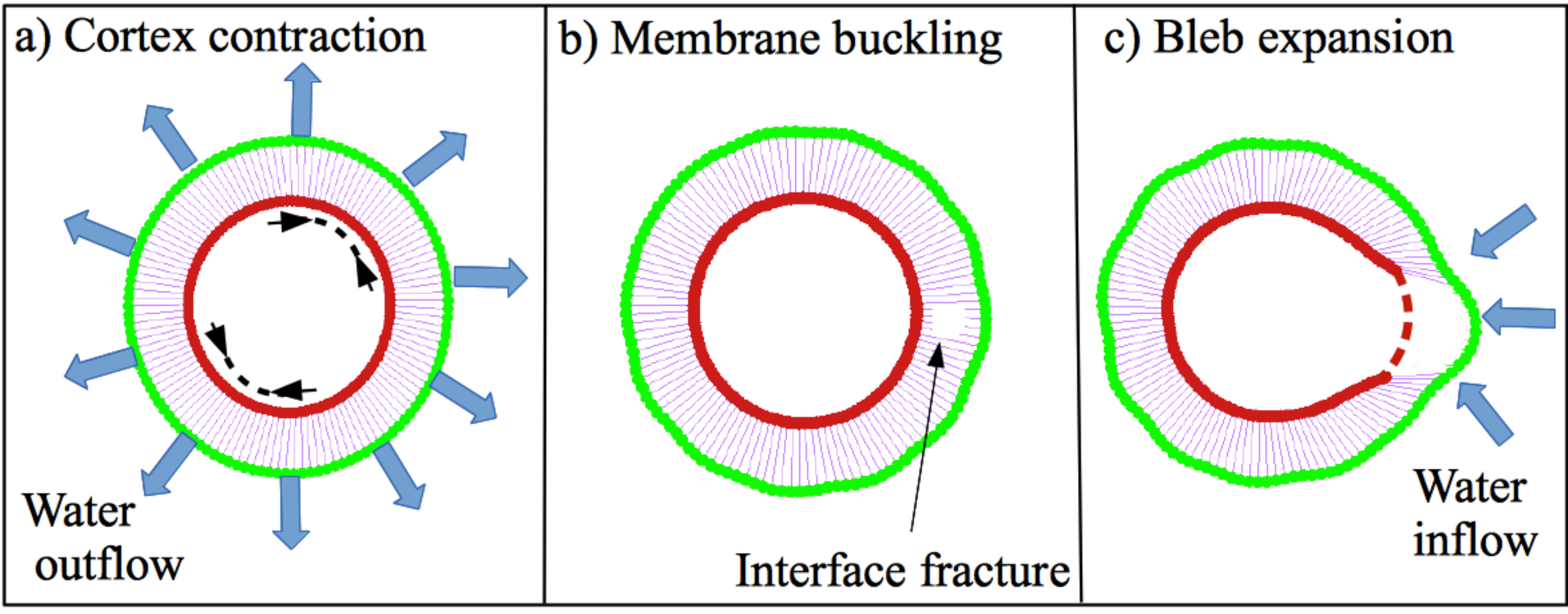}
 \caption{\label{fig:cartoon}  A schematic representation of the bleb formation process. a) The cortex contracts squeezing water outside of the cell. b) The membrane buckles and the cortex-membrane interface fractures. c) The bleb expands as the 
 interface fails and water flows inside the cell as the internal fluid pressure is relieved.}
 \end{figure}

\clearpage

\section{Supplemental tables}

\begin{table}[h]

\centering 
\begin{tabular}{|l |c c c c|}
\hline
 & WT & DNROK & AQP- & AQP+ \\ \hline
p-value ($\Delta V/\bar{V}$) &  0.11 & 0.21 &  0.17 &  0.99 \\
p-value ($\Delta\Sigma/\bar{\Sigma}$) &  0.68 & 0.75 & 0.7  & 0.13 \\
\hline
 \end{tabular}
 \caption{Results of statistical significance tests for validity of Gaussian
statistics for volume and surface fluctuations. We report the p-values
obtained from the Kolmogorov-Smirnov test. A small p-value (e.g. $p<0.01$) 
would imply that we can reject the hypothesis that the distribution is described by
Gaussian statistics. In the present case, the p-value is large indicating that
a Guassian distribution provides a good fit to the data. \label{table:pvalue}}
 
 \end{table}

\begin{table}[ht]
\centering 
\begin{tabular}{|l|c c c c c c|}
\hline
 & WT / & WT/ & WT/ & AQP-/ & AQP+/ & AQP+/ \\
  & DNROK & AQP- & AQP+ & DNROK & AQP- & DNROK \\\hline
p-value ($\Delta V/\bar{V}$) &  $5 \cdot 10^{-5}$ & $4 \cdot 10^{-4}$ & 0.25   & 0.62 &   $8 \cdot 10^{-5}$ & $8 \cdot 10^{-5}$\\
p-value ($\Delta\Sigma/\bar{\Sigma}$) & 0.001  &  0.008 & 0.09 & 0.98 &$6 \cdot 10^{-6}$ & $6 \cdot 10^{-6}$\\
\hline
 \end{tabular}
 
\caption{Results of statistical significance tests for the comparison between WT,
DNROK,AQP+,AQP- cells in the case of volume and surface
distributions. We report the p-values
obtained from the Kolmogorov-Smirnov test. A low p-value (e.g. $p<0.01$) 
indicates that we can reject the hypothesis that the two data sets are described by the
same distribution. \label{table:pvalue2}}
 \end{table}

\begin{table}[ht]
\centering 
\begin{tabular}{l c c r}
\hline\hline                        
Symbol & Quantity & Value & Reference \\ [0.5ex]
\hline                 
 $r_\mathrm{m}$& membrane radius & $25$ $\unit{\mu m}$ & \cite{Tinevez2009} \\
 $r_\mathrm{c}$& cortex radius & $24$ $\unit{\mu m}$ & \cite{Tinevez2009}  \\
 $\epsilon_\mathrm{c}=\epsilon_\mathrm{m}$ & rest length of cortex/membrane elements & $1.3$ $\unit{\mu m}$ & ---  \\
 $\Delta $ & regularization parameter & $0.5\times\epsilon$ & \cite{Cortez2001}  \\
 $k_\mathrm{m}$& membrane stiffness coefficient & $6\times 10^{-6}$ $\unit{N m^{-1}}$ &  \cite{Charras2008}\\
 $k_\mathrm{c}$& cortex stiffness coefficient & $9\times 10^{-5}$ & \cite{Charras2008}  \\
$B_\mathrm{m}$& membrane flexural rigidity & $4\times 10^{-20}$ $\unit{J}$ & \cite{Charras2008}  \\
$B_\mathrm{c}$& cortex flexural rigidity &  $2.8\times 10^{-19}$ $\unit{J}$ & \cite{Charras2008}  \\
 $k_{\mathrm{mc}}$ & cortex-membrane interface stiffness & $25\times 10^{-6}$ $\unit{N m^{-3}}$ & \cite{Charras2008} \\
 $\mu_\mathrm{c}$ & cortex drag coefficient &  $10^{-7}$  $\unit{kg m^{-2}s^{-1}}$ & \cite{Strychalski2013}  \\ 
 $\nu_\mathrm{c}$ & cortex healing speed &  $6 \times 10^{-4}$  $\unit{m s^{-1}}$ & \cite{Lim2013}  \\ 
 $\mu$ & Cytosolic viscosity   &  $10^{-1}$  $\unit{kgm^{-1}s^{-1}}$ & \cite{Lim2013}  \\
  $\alpha$ & membrane permeability  &  $10^{-4}-10^{-1}\times 10^{-12}$  $\unit{m^{2}skg^{-1}}$ & \cite{Hagedorn1997}  \\
 [1ex]      
 \hline
 \end{tabular}
 \caption{Parameters employed in numerical simulations}
 \label{table:nonlin}
 \end{table} 

\clearpage

\section{Supplemental movie captions}
\begin{itemize}
\item[\bf Movie S1] A representative example of the time evolution of a PGC in WT conditions. The movie is obtained using an average intensity 3D projection in imageJ. 
\item[\bf Movie S2]  A representative example of the time evolution of a PGC in DNROK conditions. The movie is obtained using an average intensity 3D projection in imageJ. 
\item[\bf Movie S3]  A representative example of the time evolution of a PGC in AQP- conditions. The movie is obtained using an average intensity 3D projection in imageJ. 
\item[\bf Movie S4]  A simulation of the computational model using a  porous membrane. The color represents fluid pressure (see Fig. 3a).
\item[\bf Movie S5]  A simulation of the computational model using an impermeable. The color represents fluid pressure (see Fig. 3b).
 \end{itemize}

 \end{document}